\documentclass[preprint,aps,showpacs,draft]{revtex4}

\newcommand{\sr}{\scriptstyle}
\newcommand{\ee}{{\mathrm{e}}}
\newcommand{\ii}{{\mathrm{i}}}

\usepackage{graphicx}
\usepackage{dcolumn}
\usepackage{amsmath}

\begin{document}


\title{Numerical study of surface-induced reorientation \\ and smectic
layering in a nematic liquid crystal}

\author{Joachim Stelzer}
 \email{js69190@gmx.de} 
\affiliation{Schwetzinger Str.~20, 69190 Walldorf (Baden), Germany}

\author{Ralf Bernhard}
\affiliation{IMPACT Messtechnik GmbH, 71332 Waiblingen, Germany}



\begin{abstract}
Surface-induced profiles of both nematic and smectic order parameters
in a nematic liquid crystal, ranging from an orienting substrate to
``infinity'', were evaluated numerically on base of an extended
Landau theory.  In order to obtain a smooth behavior of the solutions
at ``infinity'' a boundary energy functional was derived by
linearizing the Landau energy around its equilibrium solutions. We
find that the intrinsic wave number of the smectic structure, which
plays the r\^ole of a coupling between nematic and smectic order,
strongly influences the director reorientation. Whereas the smectic
order is rapidly decaying when moving away from the surface, the
uniaxial nematic order parameter shows an oscillatory behavior close
to the substrate, accompanied by a non-zero local biaxiality.
\end{abstract}


\maketitle


\section{INTRODUCTION}

The structure of uniaxial nematic liquid crystals is strongly
influenced by the presence of orienting surfaces \cite{Jerome}. In
addition to the usual elastic distortions in the nematic bulk, which
are well described by the Oseen-Z\"ocher-Frank energy, strong
deformations can occur close to the surface substrate
\cite{Oldano,Barbero1}.  These effects are often accompanied by a
non-zero local biaxiality of the orientational order
\cite{Poniewierski}. The nematic orientation close to a confining
substrate could be detected experimentally using second-harmonic
generation techniques \cite{Zhuang}.  In addition, by X-ray studies
the surface has also been proven to induce a layered structure, {\em
i.e.}, smectic order appears close to the surface which decays when
moving away from the substrate into the nematic bulk \cite{Als}. In
spite of considerable effort, both in theory
\cite{Barbero2,Skacej,Pergamenshchik,Barbero3} and computer simulation
\cite{Tjipto,Stelzer1,Stelzer2,Rio}, the full complexity of surface-induced
structural changes in nematics is far from being understood. Whereas
Monte Carlo or molecular dynamics simulations approach the problem on
the molecular level, in our contribution we take a phenomenological
viewpoint. To this aim we consider an extended Landau theory,
comparable to Ska\v{c}ej {\em et al.} \cite{Skacej}.  In addition to
that paper, we investigate not only uniaxial order, instead, both the
full alignment tensor for nematic order and, mainly, the amplitude of
the smectic layering is taken into account.  In particular, the
influence of the coupling between smectic and nematic order on the
order parameter profiles, obtained from numerical relaxation, is
investigated in detail. Whereas the boundary conditions at the surface
substrate are fixed, an additional boundary energy is derived, in
order to guarantee a smooth behavior of the profiles at ``infinity''
(far from the surface) where the volume equilibrium values of the
order parameters should be reached.

The organization of the article is as follows. In Section II
the Landau theory used in our calculations is introduced. 
(We focus on the derivation of the additional boundary energy
at ``infinity'' in the Appendix.) 
Section III indicates the numerical relaxation method employed
and presents selected results for order parameter profiles.
Finally, Section IV contains some concluding remarks.

\section{EXTENDED LANDAU THEORY FOR SURFACE-INDUCED EFFECTS}

The geometry of our system is the semi-infinite space ($z\ge0$), 
confined by a substrate surface at $z=0$ and
``infinity'' ($z=\infty$) which, in numerical practice,
means a large distance from the surface. Due to the infinite
extension of the system in $x$ and $y$ direction and the absence
of any lateral structure of the surface, we can reduce the problem to
a one-dimensional geometry, {\em i.e.}, all quantities only
depend on the distance $z$ from the surface. In order to
be able to investigate both positional and orientational order,
we need two different order parameters. The smectic order parameter
usually is a complex number, $\rho = \psi\,{\ee^{\ii\,\chi}}$,
whose phase $\chi$ accounts for local layer deformations. 
We assume perfect layering at the surface ($\rho(z=0) = 1$) and,
therefore, we are left only with the amplitude $\psi(z)$ 
of the layering which is a real quantity indicating 
the degree of smectic order.
The nematic order parameter is a second-rank traceless and symmetric
tensor. Without loss of generality we choose its parametrization as
\begin{equation}
{\bf{Q}}(z) =  \left(
\begin{array}{ccc}
Q_{xx}(z) & Q_{xy}(z) & Q_{xz}(z) \\ 
Q_{xy}(z) & Q_{yy}(z) & Q_{yz}(z) \\ 
Q_{xz}(z) & Q_{yz}(z) & - Q_{xx}(z) - Q_{yy}(z)  
\end{array} \right).
\label{qtensor}
\end{equation}
Therefore, there are six scalar functions whose profiles ($z$ dependence) 
have to be determined. These profiles are found from a numerical minimization
of an energy, supplied with appropriate boundary conditions
at $z=0$ and $z= \infty$.

\subsection{Bulk energy functional}

The bulk energy functional is chosen according to an extended Landau theory.
It consists of a smectic and a nematic contribution. The smectic
energy contains a volume and an elastic contribution \cite{Millan},
\begin{equation}
{\cal F}_{\mathrm{smec}} = 
\frac{\sr 1}{\sr 2}\,\tau\,| \rho | ^2 
+ | \rho | ^4
+ \frac{\sr 1}{\sr 2}\,\kappa\,
| ({\bf{\nabla}} - {\ii}\,q_0\,\delta{\bf{n}}_{\perp})\,\rho | ^2.
\label{smectenergy1}
\end{equation}
Due to the continuity of the smectic-nematic phase transition,
the volume smectic energy is an expansion
into even powers of the smectic order parameter $\rho$.
This phase transition occurs at $\tau =0$, where
$\tau$ is a reduced temperature. 
The last expression in (\ref{smectenergy1}) is the elastic
energy due to the gradients of the layer amplitude
(with smectic elastic constant $\kappa$).
It incorporates a {\em coupling to the nematic
order}, based on local $U$(1) gauge invariance, due to the
nature of the smectic order parameter as a complex number \cite{Millan}.
Namely, local changes of the smectic order are accompanied by
transversal fluctuations of the director field ${\bf n}$,
perpendicular to the layer normal. In our simplified geometry
these fluctuations are always in the $x$-$y$ plane. The coupling
strength is given by the intrinsic wave number $q_0$ of the
smectic layering. After inserting the nematic tensor order parameter
(\ref{qtensor}) and reducing the smectic order parameter to its
amplitude, the smectic energy becomes a functional dependent
on $\psi(z)$, $\psi'$ (prime denoting derivative with respect to $z$) 
and $Q_{ij}(z)$,
\begin{equation}
{\cal F}_{\mathrm{smec}} 
= \frac{\sr 1}{\sr 2}\,\tau\,\psi^2 
+ \psi^4
+ \frac{\sr 1}{\sr 2}\,\kappa\,
\left[ \psi'\,^2 + q_0^2\,
\left( Q_{xx} + Q_{yy} + \frac{\sr 2}{\sr 3}\,
S_{\mathrm{vol}} \right)\,
\psi^2 \right].
\label{smectenergy2}
\end{equation}
\noindent
Here, $S_{\mathrm{vol}}$ is the scalar order parameter that minimizes
the volume {\em nematic} energy (see next subsection).

\noindent
The nematic energy functional also consists of a volume and an elastic part,
\begin{eqnarray}
{\cal F}_{\mathrm{nem}} &=&
\frac{\sr 1}{\sr 4}\,t\,Q_{ij}\,Q_{ji}
- \sqrt{6}\,Q_{ij}\,Q_{jk}\,Q_{ki}
+ (Q_{ij}\,Q_{ji})^2
\label{nemenergy1}
\\
& & + \frac{\sr 1}{\sr 4}\,(\partial_{i}Q_{jk})\,(\partial_{i}Q_{jk})
+ \frac{\sr 1}{\sr 4}\,k_{21}\,(\partial_{i}Q_{ik})\,(\partial_{j}Q_{jk})
+ \frac{\sr 1}{\sr 4}\,k_{31}\,(\partial_{k}Q_{ij})\,(\partial_{j}Q_{ik})
\nonumber
\\
&=&
\frac{\sr 1}{\sr 2}\,t\,(Q_{xx}\,^2 + Q_{xy}\,^2 + Q_{xz}\,^2
+ Q_{yy}\,^2 + Q_{yz}\,^2 + Q_{xx}\,Q_{yy} )
\label{nemenergy2}
\\
& & - 3\,\sqrt{6}\,\left( Q_{xx}\,Q_{xy}\,^2 + Q_{yy}\,Q_{xy}\,^2 
- Q_{xx}\,Q_{yy}\,^2 - Q_{yy}\,Q_{xx}\,^2 \right.
\nonumber \\
& & \left. - Q_{xx}\,Q_{yz}\,^2 - Q_{yy}\,Q_{xz}\,^2 
+ 2\,Q_{xy}\,Q_{xz}\,Q_{yz} \right)
\nonumber \\
& & + 4\,(Q_{xx}\,^2 + Q_{xy}\,^2 + Q_{xz}\,^2
+ Q_{yy}\,^2 + Q_{yz}\,^2 + Q_{xx}\,Q_{yy} )\,^2
\nonumber \\
& & + \frac{\sr 1}{\sr 2}\,(Q'_{xx}\,^2 + Q'_{xy}\,^2 + Q'_{xz}\,^2
+ Q'_{yy}\,^2 + Q'_{yz}\,^2 + Q'_{xx}\,Q'_{yy} )
\nonumber \\
& & + \frac{\sr 1}{\sr 4}\,(k_{21} + k_{31})\cdot (Q'_{xx}\,^2 + Q'_{yy}\,^2 
+ Q'_{xz}\,^2 + Q'_{yz}\,^2 + 2\,Q'_{xx}\,Q'_{yy} ).
\nonumber 
\end{eqnarray}
Unlike the smectic energy, the volume part of (\ref{nemenergy1}) 
contains a third-order expression, to describe the discontinuous
isotropic-nematic phase transition, which, in our parametrization.
occurs at $t = \frac{9}{8}$. ($t$ is again a reduced 
temperature.) For the elastic part of (\ref{nemenergy1}) there are
three independent deformation modes, similar to the Oseen-Z\"ocher-Frank
theory for elastic distortions of the director.
$k_{21}$ and $k_{31}$ denote the ratios of elastic
constants for the respective deformation modes.

\subsection{Boundary conditions}

At the substrate surface the values for the order parameters are fixed.
We assume ideal smectic and uniaxial nematic order. That means the
smectic amplitude is one, and the alignment tensor is completely determined
by the uniaxial scalar order parameter $S=1$ and the fixed director
surface tilt angle $\Theta_{\mathrm{surf}}$ (measured 
in the $x$-$z$ plane, from the $z$ axis). This results in the
Dirichlet boundary conditions,

\begin{eqnarray}
\psi(z = 0) &=& 1 \enspace ,
\label{dirichlet1}
\\
{\bf{Q}}(z = 0) &=& S\,\left( 
{\bf{n}}\otimes{\bf{n}} - \frac{\sr 1}{\sr 3}\,
{\bf{1}}\right)
\label{quniax}
\\
&=& \left( \begin{array}{ccc}
\sin^2\Theta_{\mathrm{surf}} - \frac{\sr 1}{\sr 3} & 0 &
\sin\Theta_{\mathrm{surf}}\,\cos\Theta_{\mathrm{surf}} 
\\
0 & - \frac{\sr 1}{\sr 3} & 0 
\\
\sin\Theta_{\mathrm{surf}}\,\cos\Theta_{\mathrm{surf}} &
0 & \cos^2\Theta_{\mathrm{surf}} - \frac{\sr 1}{\sr 3} 
\label{dirichlet2}
\end{array} \right)
\end{eqnarray}

At infinity, the boundary conditions are not of Dirichlet-type.
Instead, we have to guarantee smooth profiles (zero slope)
for the order parameters which should reach
those values that minimize the {\em volume} parts of the smectic and
nematic energy (\ref{smectenergy2}) and (\ref{nemenergy2}). To this
aim we insert the uniaxial form of the alignment tensor (\ref{quniax})
into (\ref{nemenergy2}). A direct minimization yields the temperature
dependence of the volume order parameters,
\begin{eqnarray}
\psi_{\mathrm{vol}}(\tau) &=& \frac{1}{2}\,\sqrt{-\tau} \enspace ,
\label{psivol} 
\\
S_{\mathrm{vol}}(t) &=& \frac{ 3\,\sqrt{6} +
\sqrt{ 54 - 48\,t}}{16}
\label{Svol}
\end{eqnarray}

The tilt angle to be reached at infinity, $\Theta_{\infty}$, is
unknown {\em a priori}. (Previous molecular dynamics simulations
indicate a tilt angle of $\Theta_{\infty}=0$ at infinity
\cite{Stelzer1}.) We, therefore, take $\Theta_{\infty}$ as a free
parameter in our calculations. Its actual value is determined
by performing a series of simulations for any fixed set of 
the remaining simulation parameters. Monitoring $\Theta_{\infty}$
versus the energy $E_{\mathrm{equi}}$ of the equilibrated configurations
for this series yields the tilt angle at infinity, which corresponds
to the minimum of the function $\Theta_{\infty}(E_{\mathrm{equi}})$.
Following this procedure, we are able to determine 
$\Theta_{\infty}$ in dependence of the reduced temperature $t$
and the smectic wave number $q_0$.

Next we have to establish a boundary energy functional 
${\cal F}_{\mathrm{\infty}}$ whose minimization
leads to the desired surface profiles at infinity. Based upon
the procedure introduced by Galatola {\em et al.} \cite{Galatola}
the central idea for finding the boundary energy functional is a 
{\em linearization of 
the bulk equations around the volume order parameters} calculated above.
We provide all details of this derivation, which is an important ingredient 
of our method, in the Appendix. Here we merely quote the result,
\begin{eqnarray}
{\cal F}_{\infty} &=& \frac{\sr 1}{\sr 2}\,U\,( Q_{xx}\,^2 + Q_{yy}\,^2 )
+ \frac{\sr 1}{\sr 3}\,(U + V)\,S_{\mathrm{vol}}\,
(Q_{xx} + Q_{yy})
\label{boundenergy}
\\
& & - S_{\mathrm{vol}}\,\sin^2\Theta_{\infty}\cdot
(U\,Q_{xx} + V\,Q_{yy})
+ V\,Q_{xx}\,Q_{yy} + \frac{\sr 1}{\sr 2}\,\sqrt{t}\cdot 
Q_{xy}\,^2
\nonumber \\
& & + \frac{\sr 1}{\sr 2}\,
\sqrt{\frac{\sr 1}{\sr 2}\,(2 + k_{21} + k_{31})\,t}\cdot 
(Q_{xz}\,^2 
- 2\,S_{\mathrm{vol}}\,\sin\Theta_{\infty}\,\cos\Theta_{\infty}\cdot Q_{xz}
+ Q_{yz}\,^2)
\nonumber \\
& & + \frac{\sr 1}{\sr 2}\,\sqrt{\kappa\tau}\cdot\psi^2,
\nonumber
\end{eqnarray}
where $U$ and $V$ are terms which depend
on the reduced temperatures $\tau$ and $t$, the nematic
elastic constants $k_{21}$, $k_{31}$ and the volume order parameter 
$S_{\mathrm{vol}}(t)$. (For the explicit expressions see Appendix A.)

Now all order parameter profiles can be obtained from a minimization
of the total energy 
\begin{equation}
F = \int_0^\infty\,({\cal F} + {\cal F}_\infty\,
\delta(\infty))\,\mbox{d}z \enspace ,
\label{totalenergy}
\end{equation}
with the Dirichlet boundary conditions (\ref{dirichlet1}) and
(\ref{dirichlet2}) valid at $z=0$.

\section{ORDER PARAMETER PROFILES}

The minimization of the total energy (\ref{totalenergy}) was
performed numerically, employing a standard Newton-Gau{\ss}-Seidel
technique which, in our case, formally corresponds to a one-dimensional version
of the Finite Element method. First the ``infinite'' distance 
from the surface was replaced by a large, finite value 
$z_{\mathrm{max}} = 100$. The
range $0 \le z \le z_{\mathrm{max}}$ was discretized in $N = 1000$ intervals.
The bulk and boundary energy functionals (\ref{smectenergy2}),
(\ref{nemenergy2}), (\ref{boundenergy}) were evaluated on 
these intervals, the derivatives with respect to $z$ being replaced
by finite differences. The values of the order parameters
at $z=0$ were fixed according to (\ref{dirichlet1}) and (\ref{dirichlet2}).
For the initial configuration we assumed linear profiles for all
quantities $\psi$ and $Q_{ij}$ on the interval $0 \le z \le z_{\mathrm{max}}$,
by interpolating between their surface and volume values. An iterative
procedure was then performed on each grid point. All order parameters
were corrected according to the Newton-Gau{\ss}-Seidel prescription,
\begin{equation}
X^{\mathrm{new}} = X^{\mathrm{old}}
- \frac{\partial F / \partial X}{\partial^2 F / \partial X^2} 
\enspace ,
\enspace (X = \psi,\,Q_{ij})
\enspace ,
\label{NGS}
\end{equation}
where the functional derivatives in (\ref{NGS}) were evaluated 
by numerical differentiation.
The relaxation was terminated when the relative change of the total energy
was less than $10^{-6}$ which corresponded to some thousand relaxation steps.

For further discussion the nematic tensor order parameter will be
analyzed in terms of its eigenvalues and eigenvectors. More specifically,
we will plot the tilt angle of the main director, measured from the
$z$ axis and two scalar order parameters. The latter ones measure
the degree of uniaxial and biaxial order, respectively. 
(We checked that the director twist angle stays constant, due to the
surface anchoring in the $x$-$z$ plane.)

The equilibrium profiles were evaluated for different values of the
reduced temperature $t = 0 ... 1$ and the intrinsic smectic wave
number $q_0 = 0.3 ... 0.8$. All other parameters were fixed,
except for the tilt angle at infinity $\Theta_{\infty}$, which
was found for each set $(t,\,q_0)$ from the additional
minimization procedure indicated in Section II.B. 
Thereby, $\Theta_{\infty}$ could be determined up to
a maximum error of $\pm 1^{\circ}$. Regarding the remaining parameters,
the elastic constants were chosen as $k_{21} = k_{31} = 1$ and
$\kappa = 5$ which accounts for the fact that layer distortions should
contain a higher elastic energy than deformations of orientational
order. The reduced temperature was $\tau = 0.1$ which corresponds to a
nematic state point, slightly above the smectic-nematic phase
transition.  Finally, the director at the surface was anchored at a
tilt angle of $\Theta_{\mathrm{surf}}= 60^\circ$.

Let us first discuss the behavior of the tilt angle at infinity
$\Theta_{\infty}$ in dependence on the reduced temperature $t$ and the
smectic wave number $q_0$.  The respective results are displayed in
Table I. Fixing the surface tilt angle at $\Theta_{\mathrm{surf}} =
60^{\circ}$, in almost all cases we find a strong reorientation
towards the surface normal which increases with the smectic wave
number.  The influence of the smectic wave number on the director
reorientation can be understood from the particular form of the
coupling energy (\ref{smectenergy1}). In order to minimize this
coupling the transversal director components should be small in those
regions where the smectic order parameter is significantly non-zero,
{\em i.e.}, close to the surface. Therefore, it is obvious that for
increasing wave number the reorientation of the tilt angle towards the
direction of the surface normal becomes more pronounced. 
Unlike for the smectic wave number, the dependence of $\Theta_{\infty}$
on the reduced temperature $t$ is fairly small. Only for low
values of both $t$ and $q_0$ the tilt angle deformation is reduced.
Apparently, at $t=0.0$ the orientational fluctuations are still too small
to enhance the director reorientation for small wave numbers.

Figures 1-3 correspond to a reduced temperature of $t = 0.0$ which
means a nematic state point far away from the nematic-isotropic phase
transition.  The smectic order parameter profile is given in Fig.~1
for wave numbers $q_0 = 0.3$, 0.6, 0.8, corresponding to a layer
spacing of 20.9, 12.6, 7.8. Obviously, the smectic structure is
rapidly decaying from its maximum value of one when moving away from
the surface.  At a distance of 20 the curves essentially have reached
their asymptotic value of zero. This loss of smectic order is almost
independent of the wave number.

As shown in Figure 2 the uniaxial nematic order parameter is also
approaching its asymptotic value within a distance of 20. The value
reached is 0.92, which precisely corresponds to the volume value of
the scalar order parameter at temperature $t=0.0$, according to
(\ref{Svol}).  Remarkably, the uniaxial order parameter is not
decaying monotonically, instead there is an oscillatory behavior. For
large wave numbers ($q\ge 0.6$) it even decreases below its volume
value. The non-monotonic behavior of uniaxial order is accompanied by
the occurence of a non-zero biaxial order parameter.  Both suppression
of uniaxial order and increased biaxiality close to the surface have
also been observed in computer simulations based on the molecular
Gay-Berne model \cite{Stelzer1}.

Unlike the scalar order parameters, the profile of the director tilt
angle strongly depends on the intrinsic smectic wave number (Figure
3).  Whereas for $q_0=0.8$ there is again a strong change within a
comparably short distance, for low wave numbers ($q \le 0.6$) the
director reorientation is much weaker. The behavior of the tilt angle
is also changing qualitatively with the wave number. {\em E.g.}, for
$q_0 = 0.8$ the tilt angle profile becomes non-monotonous, taking
intermediate values that are closer to the homeotropic orientation
than the tilt angle $\Theta_{\infty}$ finally reached.  As revealed
from Figure 4, an even more drastic change occurs for the reduced
temperature $t = 1.0$, which is just below the nematic-isotropic phase
transition.  For low wave numbers the tilt angle profiles show a local
maximum at a distance of around 20, before decaying towards the volume
value.

\section{REMARKS}

\begin{enumerate}

\item Summarizing our work, we have numerically analyzed the
sur\-face-\-in\-du\-ced profiles of smectic and nematic order as well
as director orientation. Whereas the order parameters are always
strongly changing in a thin layer close to the surface, the tilt angle
reorientation is mainly dependent on the intrinsic smectic wave number
of the liquid crystal which, in our model, acts as a coupling
parameter between nematic and smectic order. It seems, however, that
the case of high wave number ($q=0.8$) is the most realistic,
considering the experimental observation of a strong reorientation
close to the surface \cite{Zhuang}. In addition, surface-induced
biaxiality and suppression of uniaxial order, previously detected in
experiment and molecular simulations, could also be confirmed within
the frame of our model.

\item The asymptotic behavior of the profiles at infinity was
reproduced in numerics by deriving a boundary energy functional.
To this aim, referring to the method of Ref.~\cite{Galatola} 
we linearized the bulk Euler-Lagrange equations around their
volume solutions. When the boundary energy is included into the
numerical minimization procedure, the order parameter profiles at
infinity smoothly reach their volume values. Although in our case
this technique was used in a one-dimensional geometry, it could be
extended in a straightforward way to more complex situations.  For
instance, a laterally structured surface, which is fairly common in
novel developments of display technique, will already break the
symmetry of our example. However, the treatment of open
boundaries by an additional energy functional derived from a
linearization of the bulk energy is a very general concept which could
be used for any numerical study dealing with ``infinitely'' extended
systems.

\end{enumerate}

\appendix

\section{DERIVATION OF THE BOUNDARY ENERGY FUNCTIONAL}

Before starting to derive the boundary energy functional at infinity, 
we first remind of the desired values for the order parameters.
At infinity the smectic order is completely lost ($\psi(z=\infty) = 0$). 
The nematic order parameter is again uniaxial, but with the
volume scalar order parameter $S_{\mathrm{vol}}(t)$, dependent on temperature, 
and tilt angle $\Theta_{\infty}$,
\begin{eqnarray}
\psi(z = \infty) &=& 0
\label{psiinfty}
\\
{\bf Q}(z = \infty) &=& S\,\left( {\bf n}\otimes{\bf n} - \frac{\sr 1}{\sr 3}\,
{\bf 1}\right)
\nonumber \\
&=& \left( \begin{array}{ccc}
S_{\mathrm{vol}}(t)\,\left(\sin^2\Theta_{\infty} - \frac{\sr 1}{\sr 3}\right)
& 0 & S_{\mathrm{vol}}(t)\,\sin\Theta_{\infty}\,\sin\Theta_{\infty} \\
0 & - \frac{\sr 1}{\sr 3}\,S_{\mathrm{vol}}(t) & 0 \\
S_{\mathrm{vol}}(t)\,\sin\Theta_{\infty}\,\sin\Theta_{\infty} &
0 & S_{\mathrm{vol}}(t)\,\left(\cos^2\Theta_{\infty} - \frac{\sr 1}{\sr 3}
\right)
\end{array} \right)
\label{qtensorinfty}
\end{eqnarray}

We start with the bulk equations which are obtained from the bulk
energy functional ${\cal F} = {\cal F}_{\mathrm{smec}} + {\cal
F}_{\mathrm{nem}}$, (\ref{smectenergy2}) and (\ref{nemenergy2}), by
variational calculus as the corresponding Euler-Lagrange equations,
\begin{equation}
L_{X} = \frac{\partial{\cal F}}{\partial X}
- \frac{\mathrm{d}}{\mathrm{d}z}\,
\frac{\partial{\cal F}}{\partial X'} = 0 \enspace , \enspace
(X = \psi,\,Q_{ij}).
\label{eulerlagrange}
\end{equation}
In order to obtain a smooth behavior of the profiles at infinity, the
Euler-Lagrange equations (\ref{eulerlagrange}) are linearized around
the volume order parameters. To that aim we calculate the partial
derivatives of the right-hand sides at infinity which, in turn,
are obtained from the quadratic part of the free energy
(\ref{smectenergy2}) and (\ref{nemenergy2}),
\begin{eqnarray}
\left.\frac{\partial L_{xx}}{\partial Q_{xx}}\right|_{z = \infty}
= \left.\frac{\partial L_{yy}}{\partial Q_{yy}}\right|_{z = \infty}
= \left.\frac{\partial L_{xy}}{\partial Q_{xy}}\right|_{z = \infty}
= \left.\frac{\partial L_{xz}}{\partial Q_{xz}}\right|_{z = \infty}
= \left.\frac{\partial L_{yz}}{\partial Q_{yz}}\right|_{z = \infty} 
&=& t \enspace ,
\\
\left.\frac{\partial L_{xx}}{\partial Q_{yy}}\right|_{z = \infty}
= \left.\frac{\partial L_{yy}}{\partial Q_{xx}}\right|_{z = \infty} 
&=& \frac{1}{2}\,t \enspace ,
\\
\left.\frac{\partial L_{\psi}}{\partial \psi}\right|_{z = \infty}
&=& \tau. 
\end{eqnarray}
The linearized Euler-Lagrange equations now become explicitly
\begin{eqnarray}
Q''_{xx} &=& \frac{(3 + k_{21} + k_{31})\,t}{3 + 2\,(k_{21} + k_{31})}\,
\left( Q_{xx} 
- S_{\mathrm{vol}}\,\left(\sin^2\Theta_{\infty} - \frac{\sr 1}{\sr 3}
\right) \right)
\label{qxxlinear}
\\
& & - \frac{(k_{21} + k_{31})\,t}{3 + 2\,(k_{21} + k_{31})}\,
\left( Q_{yy} + \frac{\sr 1}{\sr 3}\,S_{\mathrm{vol}} \right) \enspace ,
\nonumber \\ & & \nonumber \\
Q''_{yy} &=& - \frac{(k_{21} + k_{31})\,t}{3 + 2\,(k_{21} + k_{31})}\,
\left( Q_{xx} 
- S_{\mathrm{vol}}\,\left(\sin^2\Theta_{\infty} - \frac{\sr 1}{\sr 3}
\right) \right)
\label{qyylinear}
\\ 
& & + \frac{(3 + k_{21} + k_{31})\,t}{3 + 2\,(k_{21} + k_{31})}\,
\left( Q_{yy} + \frac{\sr 1}{\sr 3}\,S_{\mathrm{vol}} \right) \enspace ,
\nonumber \\ & & \nonumber \\
Q''_{xy} &=& t\,Q_{xy} \enspace ,
\label{qxylinear}
\\ & & \nonumber \\
Q''_{xz} &=& \frac{2\,t}{2 + k_{21} + k_{31}}\,
\left( Q_{xz}  
- S_{\mathrm{vol}}\,\sin\Theta_{\infty}\,\sin\Theta_{\infty}\right) 
 \enspace ,
\label{qxzlinear}
\\ & & \nonumber \\
Q''_{yz} &=& \frac{2\,t}{2 + k_{21} + k_{31}}\,Q_{yz} \enspace ,
\label{qyzlinear}
\\ & & \nonumber \\
\psi'' &=& 
\frac{\tau}{\kappa}\,\psi.
\label{psilinear}
\end{eqnarray}

Among the equations (\ref{qxxlinear})--(\ref{psilinear}), the first
two are coupled. In order to find their solutions we note that they
can be written in matrix form,
\begin{equation}
\left( \begin{array}{c}
Q''_{xx} \\ Q''_{yy} \end{array} \right)
= \left( \begin{array}{cc}
A & B \\ B & A \end{array} \right)\cdot
\left( \begin{array}{c}
Q_{xx} 
- S_{\mathrm{vol}}\,\left(\sin^2\Theta_{\infty} - \frac{\sr 1}{\sr 3}\right)
\\
Q_{yy} + \frac{\sr 1}{\sr 3}\,S_{\mathrm{vol}} 
\end{array} \right).
\end{equation}
Now we change the variables to the deviations of
$Q_{xx}$ and $Q_{yy}$ from the volume solutions,
\begin{equation}
\left( \begin{array}{c}
\delta Q''_{xx} \\ \delta Q''_{yy} \end{array} \right)
= \left( \begin{array}{cc}
A & B \\ B & A \end{array} \right)\cdot
\left( \begin{array}{c}
\delta Q_{xx} \\
\delta Q_{yy} 
\end{array} \right).
\end{equation}
By determining
the eigenvalues and eigenvectors of the coefficient matrix, it can be
expressed by its diagonalized form and the corresponding
orthogonal transformations,
\begin{equation}
\left( \begin{array}{c}
\delta Q''_{xx} \\ \delta Q''_{yy} \end{array} \right)
= \frac{\sr 1}{\sr \sqrt{2}}\,\left( \begin{array}{cc}
1 & 1 \\ - 1 & 1 \end{array} \right)\cdot
\left( \begin{array}{cc}
A - B & 0 \\ 0 & A + B \end{array} \right)\cdot
\frac{\sr 1}{\sr \sqrt{2}}\,\left( \begin{array}{cc}
1 & - 1 \\ 1 & 1 \end{array} \right)\cdot
\left( \begin{array}{c}
\delta Q_{xx} \\ \delta Q_{yy}
\end{array} \right).
\end{equation}
Now the system can be decoupled by a similarity transformation.
Keeping only the decaying modes, we arrive at the solution
\begin{equation}
\left( \begin{array}{c}
\delta Q_{xx} \\ \delta Q_{yy} \end{array} \right)
= \frac{\sr 1}{\sr \sqrt{2}}\,\left( \begin{array}{cc}
1 & 1 \\ - 1 & 1 \end{array} \right)\cdot
\left( \begin{array}{cc}
\ee^{- \sqrt{A - B}\,z} & 0 \\ 
0 & \ee^{- \sqrt{A + B}\,z} \end{array} \right)\cdot
\frac{\sr 1}{\sr \sqrt{2}}\,\left( \begin{array}{cc}
1 & - 1 \\ 1 & 1 \end{array} \right)\cdot
\left( \begin{array}{c}
\alpha \\ \beta
\end{array} \right).
\end{equation}
The asymptotic Cauchy boundary conditions are obtained
by differentiating the solutions by $z$, taken at infinity,
\begin{equation}
\left( \begin{array}{c}
\delta Q'_{xx} \\ \delta Q'_{yy} \end{array} \right)
= - \frac{\sr 1}{\sr \sqrt{2}}\,\left( \begin{array}{cc}
1 & 1 \\ - 1 & 1 \end{array} \right)\cdot
\left( \begin{array}{cc}
\sqrt{A - B} & 0 \\ 
0 & \sqrt{A + B} \end{array} \right)\cdot
\frac{\sr 1}{\sr \sqrt{2}}\,\left( \begin{array}{cc}
1 & - 1 \\ 1 & 1 \end{array} \right)\cdot
\left( \begin{array}{c}
\delta Q_{xx} \\ \delta Q_{yy}
\end{array} \right).
\end{equation}
The explicit boundary conditions for $Q_{xx}$ and $Q_{yy}$ then are
\begin{eqnarray}
Q'_{xx}(z = \infty) &=&  - \frac{\sr 1}{\sr 2}\,
\bigg( \sqrt{A + B} + \sqrt{A - B} \bigg)\cdot
\left( Q_{xx} 
- S_{\mathrm{vol}}\,\left(\sin^2\Theta_{\infty} - \frac{\sr 1}{\sr 3}\right)
\right)
\label{qxxcauchy}
\\
& &  - \frac{\sr 1}{\sr 2}\,
\bigg( \sqrt{A + B} - \sqrt{A - B} \bigg)\cdot
\left( Q_{yy} + \frac{\sr 1}{\sr 3}\,S_{\mathrm{vol}} \right) \enspace , 
\nonumber \\ & & \nonumber \\
Q'_{yy}(z = \infty) &=&  - \frac{\sr 1}{\sr 2}\,
\bigg( \sqrt{A + B} - \sqrt{A - B} \bigg)\cdot
\left( Q_{xx} 
- S_{\mathrm{vol}}\,\left(\sin^2\Theta_{\infty} - \frac{\sr 1}{\sr 3}\right)
\right)
\label{qyycauchy}
\\
& &  - \frac{\sr 1}{\sr 2}\,
\bigg( \sqrt{A + B} + \sqrt{A - B} \bigg)\cdot
\left( Q_{yy} + \frac{\sr 1}{\sr 3}\,S_{\mathrm{vol}} \right).
\nonumber
\end{eqnarray}
The remaining linearized Euler-Lagrange equations are already
decoupled which immediately yields the corresponding
Cauchy boundary conditions at infinity,
\begin{eqnarray}
Q'_{xy}(z = \infty) &=& - \sqrt{t}\cdot Q_{xy} \enspace , 
\label{qxycauchy}
\\ & & \nonumber \\
Q'_{xz}(z = \infty) &=& 
- \sqrt{\frac{2\,t}{2 + k_{21} + k_{31}}}\cdot\left( Q_{xz}
- S_{\mathrm{vol}}\,\sin\Theta_{\infty}\,\sin\Theta_{\infty}\right) 
\label{qxzcauchy}
\enspace , 
\\ & & \nonumber \\
Q'_{yz}(z = \infty) &=& 
- \sqrt{\frac{2\,t}{2 + k_{21} + k_{31}}}\cdot Q_{yz}
\label{qyzcauchy}
\enspace , 
\\ & & \nonumber \\
\psi'(z = \infty) &=& 
- \sqrt{\frac{\tau}{\kappa}}\cdot\psi.
\label{psicauchy}
\end{eqnarray}

The Cauchy boundary condition at infinity has to be derived 
from variational calculus in order to be included into
the relaxation. It is obtained from the
bulk energy functional ${\cal F}$ and a boundary energy functional
${\cal F}_{\infty}$,
\begin{equation}
\frac{\partial {\cal F}_{\infty}}{\partial X}
+ \frac{\partial {\cal F}}{\partial X'} = 0 \enspace , \enspace 
(X = \psi,\,Q_{ij}).
\label{surfvar}
\end{equation}
To obtain this boundary energy functional we insert the Cauchy
boundary conditions (\ref{qxxcauchy})--(\ref{psicauchy}), previously
derived from the linearized Euler-Lagrange equations, into the
variational equation (\ref{surfvar}). This immediately yields the six
partial derivatives of the boundary energy functional,
\begin{eqnarray}
\left( \begin{array}{c}
\frac{\displaystyle \partial {\cal F}_{\infty}}{
\displaystyle \partial Q_{xx}} \\
\frac{\displaystyle \partial {\cal F}_{\infty}}{
\displaystyle \partial Q_{yy}}
\end{array} \right) &=&
- \frac{\sr 1}{\sr 2}\, 
\left( \begin{array}{cc}
2 + k_{21} + k_{31} & 1 + k_{21} + k_{31} \\
1 + k_{21} + k_{31} & 2 + k_{21} + k_{31} 
\end{array} \right)\cdot
\left( \begin{array}{c}
Q'_{xx} \\ Q'_{yy}
\end{array} \right)
\\
&=& \frac{\sr 1}{\sr 4}\, 
\left( \begin{array}{cc}
2 + k_{21} + k_{31} & 1 + k_{21} + k_{31} \\
1 + k_{21} + k_{31} & 2 + k_{21} + k_{31} 
\end{array} \right)\cdot
\\
& & \left( \begin{array}{cc}
\sqrt{A+B} + \sqrt{A-B} & \sqrt{A+B} - \sqrt{A-B} \\
\sqrt{A+B} - \sqrt{A-B} & \sqrt{A+B} + \sqrt{A-B} 
\end{array} \right)\cdot
\nonumber \\
& & \left( \begin{array}{c}
Q_{xx} 
- S_{\mathrm{vol}}\,\left(\sin^2\Theta_{\infty} - \frac{\sr 1}{\sr 3}\right)
\\
Q_{yy} + \frac{\sr 1}{\sr 3}\,S_{\mathrm{vol}} \\ 
\end{array} \right)
\nonumber \\
&=& 
\left( \begin{array}{cc}
U & V \\ V & U
\end{array} \right)\cdot
\left( \begin{array}{c}
Q_{xx} 
- S_{\mathrm{vol}}\,\left(\sin^2\Theta_{\infty} - \frac{\sr 1}{\sr 3}\right)
\\
Q_{yy} + \frac{\sr 1}{\sr 3}\,S_{\mathrm{vol}} \\ 
\end{array} \right) 
\end{eqnarray}
\begin{equation}
\frac{\displaystyle \partial {\cal F}_{\infty}}{
\displaystyle \partial Q_{xy}}
= - Q'_{xy}  = {\sqrt{t}}\cdot Q_{xy}
\end{equation}
\begin{equation}
\frac{\displaystyle \partial {\cal F}_{\infty}}{
\displaystyle \partial Q_{xz}}
= - \frac{\sr 1}{\sr 2}\,(2 + k_{21} + k_{31})\,Q'_{xz}  
= \sqrt{\frac{\sr 1}{\sr 2}\,(2 + k_{21} + k_{31})\,t}\cdot 
\left( Q_{xz}
- S_{\mathrm{vol}}\,\sin\Theta_{\infty}\,\sin\Theta_{\infty}\right) 
\end{equation}
\begin{equation}
\frac{\displaystyle \partial {\cal F}_{\infty}}{
\displaystyle \partial Q_{yz}}
= - \frac{\sr 1}{\sr 2}\,(2 + k_{21} + k_{31})\,Q'_{yz}  
= \sqrt{\frac{\sr 1}{\sr 2}\,(2 + k_{21} + k_{31})\,t}\cdot Q_{yz}
\end{equation}
\begin{eqnarray}
\frac{\displaystyle \partial {\cal F}_{\infty}}{
\displaystyle \partial \psi}
&=& - \kappa\,\psi'
\\  
&=& \sqrt{\kappa\tau}\cdot\,\psi
\end{eqnarray}
Now the boundary energy functional is obtained by integration,
just in the same way a potential field is calculated
from a given force field in classical mechanics. The final
result, quoted in the main text (\ref{boundenergy}), reads
\begin{eqnarray}
{\cal F}_{\infty} &=& \frac{\sr 1}{\sr 2}\,U\,( Q_{xx}\,^2 + Q_{yy}\,^2 )
+ \frac{\sr 1}{\sr 3}\,(U + V)\,S_{\mathrm{vol}}\,
(Q_{xx} + Q_{yy})
\\
& & - S_{\mathrm{vol}}\,\sin^2\Theta_{\infty}\cdot
(U\,Q_{xx} + V\,Q_{yy})
+ V\,Q_{xx}\,Q_{yy} + \frac{\sr 1}{\sr 2}\,\sqrt{t}\cdot 
Q_{xy}\,^2
\nonumber \\
& & + \frac{\sr 1}{\sr 2}\,
\sqrt{\frac{\sr 1}{\sr 2}\,(2 + k_{21} + k_{31})\,t}\cdot 
(Q_{xz}\,^2 
- 2\,S_{\mathrm{vol}}\,\sin\Theta_{\infty}\,\cos\Theta_{\infty}\cdot Q_{xz}
+ Q_{yz}\,^2)
\nonumber \\
& & + \frac{\sr 1}{\sr 2}\,\sqrt{\kappa\tau}\cdot\psi^2.
\nonumber
\end{eqnarray}

\newpage


\newpage

%
%
\begin{figure}
\caption{Profile of the smectic order parameter
$\psi(z)$ for reduced temperature $t = 0.0$, 
at various intrinsic smectic wave numbers $q_0$. 
Solid line: $\psi(z)$ at $q_0 = 0.3$;
dashed line: $\psi(z)$ at $q_0 = 0.6$;
dotted line: $\psi(z)$ at $q_0 = 0.8$.}
\end{figure}

%
%
\begin{figure}
\caption{Profile of the nematic order parameters
$S(z)$ and $T(z)$ for uniaxial and biaxial order,
respectively, for reduced temperature $t = 0.0$, 
at various intrinsic smectic wave numbers $q_0$. 
Solid line: $S(z)$ at $q_0 = 0.3$;
dashed line: $S(z)$ at $q_0 = 0.6$;
upper dotted line: $S(z)$ at $q_0 = 0.8$;
lower dotted line: $T(z)$ at $q_0 = 0.8$.}
\end{figure}

%
%
\begin{figure}
\caption{Profile of the director tilt angle
$\Theta(z)$ for reduced temperature $t = 0.0$, 
at various intrinsic smectic wave numbers $q_0$. 
Solid line: $\Theta(z)$ at $q_0 = 0.3$;
dashed line: $\Theta(z)$ at $q_0 = 0.6$;
dotted line: $\Theta(z)$ at $q_0 = 0.8$.}
\end{figure}

%
%
\begin{figure}
\caption{Profile of the director tilt angle
$\Theta(z)$ for reduced temperature $t = 1.0$, 
at various intrinsic smectic wave numbers $q_0$. 
Solid line: $\Theta(z)$ at $q_0 = 0.3$;
dashed line: $\Theta(z)$ at $q_0 = 0.6$;
dotted line: $\Theta(z)$ at $q_0 = 0.8$.}
\end{figure}

%
%
\begin{table}
\begin{tabular}{ccccccc}
\hline
\hline
\qquad  \qquad $q_0$ \qquad  & 
\qquad 0.3  \qquad & 
\qquad 0.4  \qquad & 
\qquad 0.5  \qquad & 
\qquad 0.6  \qquad & 
\qquad 0.7  \qquad & 
\qquad 0.8  \qquad \\
$t$ \qquad \qquad \qquad  & & & & & & \\ 
\hline 
0.0 \qquad \qquad \qquad &
\qquad $51^{\circ}$ \qquad & 
\qquad $43^{\circ}$ \qquad & 
\qquad $38^{\circ}$ \qquad & 
\qquad $33^{\circ}$ \qquad & 
\qquad $27^{\circ}$ \qquad & 
\qquad $15^{\circ}$ \qquad \\ 
0.1 \qquad \qquad \qquad &
\qquad $20^{\circ}$ \qquad & 
\qquad $20^{\circ}$ \qquad & 
\qquad $17^{\circ}$ \qquad & 
\qquad $15^{\circ}$ \qquad & 
\qquad $13^{\circ}$ \qquad & 
\qquad $8^{\circ}$ \qquad \\ 
0.2 \qquad \qquad \qquad &
\qquad $17^{\circ}$ \qquad & 
\qquad $16^{\circ}$ \qquad & 
\qquad $14^{\circ}$ \qquad & 
\qquad $14^{\circ}$ \qquad & 
\qquad $10^{\circ}$ \qquad & 
\qquad $6^{\circ}$ \qquad \\ 
0.3 \qquad \qquad \qquad &
\qquad $15^{\circ}$ \qquad & 
\qquad $15^{\circ}$ \qquad & 
\qquad $12^{\circ}$ \qquad & 
\qquad $12^{\circ}$ \qquad & 
\qquad $8^{\circ}$ \qquad & 
\qquad $3^{\circ}$ \qquad \\ 
0.4 \qquad \qquad \qquad &
\qquad $13^{\circ}$ \qquad & 
\qquad $12^{\circ}$ \qquad & 
\qquad $11^{\circ}$ \qquad & 
\qquad $10^{\circ}$ \qquad & 
\qquad $8^{\circ}$ \qquad & 
\qquad $3^{\circ}$ \qquad \\ 
0.5 \qquad \qquad \qquad &
\qquad $13^{\circ}$ \qquad & 
\qquad $11^{\circ}$ \qquad & 
\qquad $10^{\circ}$ \qquad & 
\qquad $10^{\circ}$ \qquad & 
\qquad $6^{\circ}$ \qquad & 
\qquad $1^{\circ}$ \qquad \\ 
0.6 \qquad \qquad \qquad &
\qquad $13^{\circ}$ \qquad & 
\qquad $11^{\circ}$ \qquad & 
\qquad $9^{\circ}$ \qquad & 
\qquad $9^{\circ}$ \qquad & 
\qquad $7^{\circ}$ \qquad & 
\qquad $1^{\circ}$ \qquad \\ 
0.7 \qquad \qquad \qquad &
\qquad $11^{\circ}$ \qquad & 
\qquad $11^{\circ}$ \qquad & 
\qquad $10^{\circ}$ \qquad & 
\qquad $8^{\circ}$ \qquad & 
\qquad $6^{\circ}$ \qquad & 
\qquad $2^{\circ}$ \qquad \\ 
0.8 \qquad \qquad \qquad &
\qquad $12^{\circ}$ \qquad & 
\qquad $10^{\circ}$ \qquad & 
\qquad $8^{\circ}$ \qquad & 
\qquad $7^{\circ}$ \qquad & 
\qquad $4^{\circ}$ \qquad & 
\qquad $1^{\circ}$ \qquad \\ 
0.9 \qquad \qquad \qquad &
\qquad $10^{\circ}$ \qquad & 
\qquad $10^{\circ}$ \qquad & 
\qquad $7^{\circ}$ \qquad & 
\qquad $6^{\circ}$ \qquad & 
\qquad $2^{\circ}$ \qquad & 
\qquad $1^{\circ}$ \qquad \\ 
1.0 \qquad \qquad \qquad &
\qquad $11^{\circ}$ \qquad & 
\qquad $9^{\circ}$ \qquad & 
\qquad $7^{\circ}$ \qquad & 
\qquad $5^{\circ}$ \qquad & 
\qquad $3^{\circ}$ \qquad & 
\qquad $1^{\circ}$ \qquad \\ 
\hline
\hline
\end{tabular}
\caption{Tilt angle at infinity $\Theta_{\infty}$ [degrees] in dependence
on the reduced temperature $t$ (rows) and the intrinsic smectic
wave number $q_0$ (columns). Both $t$ and $q_0$ are in reduced units.}
\end{table}

\end{document}